\documentclass[12pt,preprint]{aastex}

\begin{document}
\title{Blazar Anti-Sequence of Spectral Variation within individual Blazars: Cases for Mrk~501 and 3C~279}
\author{Jin Zhang\altaffilmark{1,2}, Shuang-Nan Zhang\altaffilmark{3,1}, En-Wei Liang\altaffilmark{4,1}}
\altaffiltext{1}{National Astronomical Observatories, Chinese Academy of Sciences, Beijing, 100012, China;
zhang.jin@hotmail.com}\altaffiltext{2}{Key Laboratory for the Structure and Evolution of Celestial Objects,
Chinese Academy of Sciences, Kunming, 650011, China} \altaffiltext{3}{Key Laboratory of Particle Astrophysics,
Institute of High Energy Physics, Chinese Academy of Sciences, Beijing, 100049,
China}\altaffiltext{4}{Department of Physics and GXU-NAOC Center for Astrophysics and Space Sciences, Guangxi
University, Nanning, 530004, China}

\begin{abstract}
The jet properties of Mrk 501 and 3C 279 are derived by fitting the broadband spectral energy distributions with the lepton models. The derived $\gamma_{\rm b}$ (the break Lorenz factor of the electron distribution) are $10^4-10^6$ for Mrk~501 and $200\sim 600$ for 3C~279 and the magnetic field strength ($B$) of Mrk~501 is usually one order of magnitude lower than that of 3C~279, but their Doppler factors ($\delta$) are comparable. A spectral variation feature that the peak luminosity is correlated with the peak frequency, which is opposite to the blazar sequence, is observed in the two sources.  We find that (1) the peak luminosities of the two bumps in SEDs for Mrk 501 depend on $\gamma_b$ in both the observer and co-moving frames, but they are not correlated with $B$ and $\delta$; (2)the luminosity variation of 3C 279 is dominated by the external Compton (EC) peak and its peak luminosity is correlated with $\gamma_b$ and $\delta$, but anti-correlated with $B$. These results suggest that $\gamma_{\rm b}$ may govern the spectral variation of Mrk 501 and $\delta$ and $B$ would be responsible for the spectral variation of 3C 279. The narrow distribution of $\gamma_{\rm b}$ and the correlation of $\gamma_{\rm b}$ and $B$ in 3C 279 would be due to the cooling by the EC process and the strong magnetic field. Based on our brief discussion, we propose that this spectral variation feature may be originated from the instability of the corona but not from the variation of the accretion rate as that for the blazar sequence.
\end{abstract}

\keywords{radiation mechanisms: non-thermal---gamma-rays: observations---gamma-rays: theory---Blazars:
individual (Mrk 501 and 3C~279)}

\section{Introduction}           
\label{sect:intro} Blazars, a sub-sample of active galactic nuclei (AGNs), are composed of BL Lac objects (BL
Lacs) and flat spectrum radio quasars (FSRQs). Their broadband spectral energy distributions (SEDs) are usually
bimodal. The bump in the IR-optical-UV band can be explained with the synchrotron process of relativistic
electrons accelerated in the jets, and the bump in the GeV-TeV gamma-ray band may be due to the inverse Compton
(IC) scattering of the same electron population. The seed photon fields for BL Lacs are thought to be the
synchrotron radiation themselves (the so-called SSC model, Maraschi et al. 1992; Ghisellini et al. 1996; Zhang
et al. 2012a), but the contributions of inverse Compton scattering by some external radiation fields (EC), such
as the broad line region (BLR), usually dominate the IC peak of the SEDs for FSRQs (Dermer et al. 1992).

Significant variability in multi-wavelengths with spectral shift is observed in blazars (Massaro et al. 2008;
Tramacere et al. 2009; Zhang et al. 2012a) and its physical reason is still uncertain (e.g., Zhang et al.
2012a). Associated flux variations in the X-ray and gamma-ray bands were observed in some sources
(Takahashi et al. 2000; Sambruna et al. 2000). Recently, Zhang et al. (2012a) analyzed the SEDs of GeV-TeV BL
Lacs and found that these SEDs are well modeled with the one zone Syn+SSC model. They showed that the observed
luminosity variations are usually accompanied with the shift of peak frequencies of the SEDs (see also Tavecchio
et al. 2001; B{\l}a{\.z}ejowski et al. 2005) and the spectral shift may be related to the electron
acceleration in the jet. This is consistent with that reported by Massaro et al. (2008), Tramacere et al.
(2009), and Fossati et al. (2008), who suggested that the correlation between the SED peak energy of synchrotron
radiation and the corresponding peak flux would be a signature of the acceleration or injection of the
relativistic electrons at the higher energy end by stochastic acceleration mechanism. However, some ``orphan"
flares detected in one band without counterparts in other bands were also observed (Krawczynski et al. 2004;
B{\l}a\.{z}ejowski et al. 2005). It is uncertain that these flares are from different radiating regions or due to the temporal evolution of the parameters for the emitting region (Fossati et al. 2008).

Well-sampled SEDs in different episodes for a given blazar are good probes for the jet properties
since the differences in both the jets and their environments among sources should be eliminated. This work dedicates to investigate the spectral variations in individual blazars using broadband SEDs with significant variation in different
episodes.  We compile the SEDs for Mrk~501 (a typical BL Lac) and 3C~279 (a typical FSRQ). The data are presented in \S 2.  They are fitted with the syn+SSC or syn+SSC+EC models. Our models and the derived parameters are
described in \S 3. The possible relations among parameters in both the observed and co-moving frames and the
corresponding implications are presented in \S 4. Discussion on the spectral variation
in an individual blazar and the blazar sequence is given in \S 5. A summary of our results is presented
in $\S$6.

\section{Data}
\label{sect:data} 
\begin{figure}
\includegraphics[angle=0,scale=0.35]{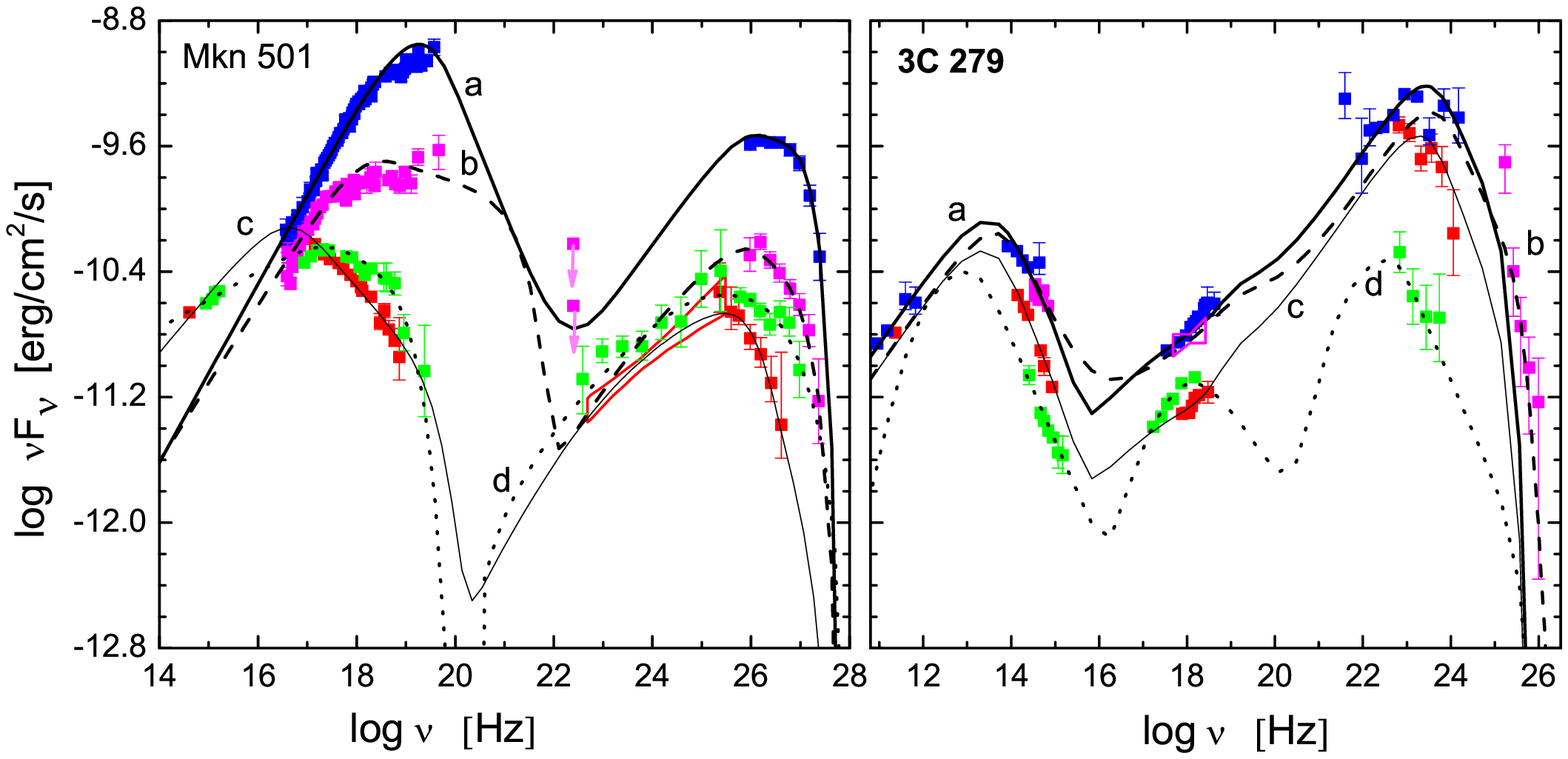}\\
\includegraphics[angle=0,scale=0.35]{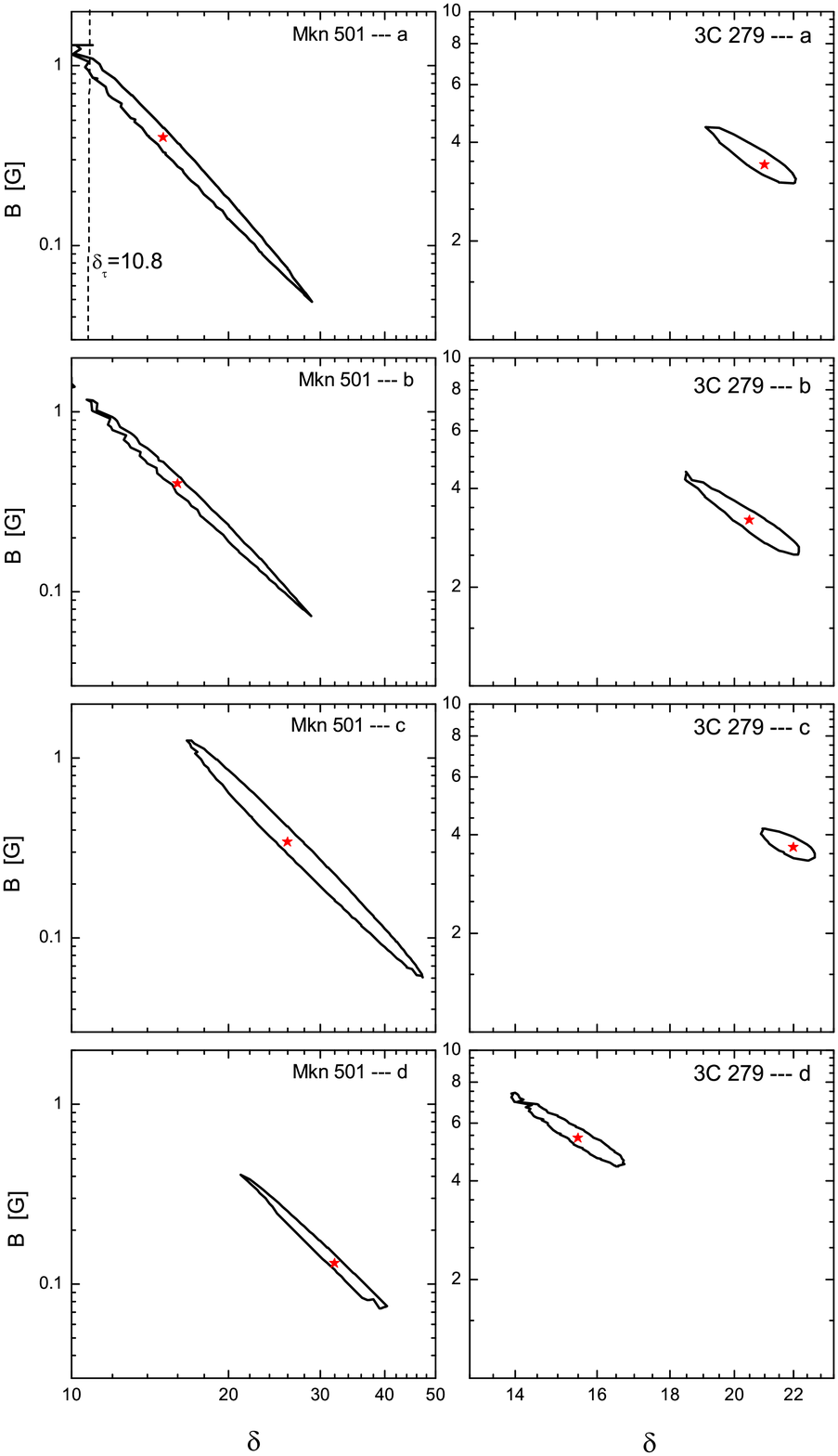}
\caption{The observed SEDs ({\em scattered data points}) with model fitting ({\em lines}) for Mrk 501 and
3C~279, and the contours of 1-$\sigma$ significance level in the $B$-$\delta$ plane. The four SEDs are named as
states ``a" ({\em blue points and thick solid line}), ``b" ({\em magenta points and dashed line}), ``c" ({\em
red points and thin solid line}), and ``d" ({\em green points and dotted line}), respectively. The red stars
stand for the best fit parameter sets of $\delta$ and $B$.}\label{SED}
\end{figure}

We compile the SEDs of Mrk 501 and FSRQ 3C~279 from literature.
We get four well-sampled SEDs for both of them as described bellow. They are shown in Figure \ref{SED}. We mark the SEDs as ``a", ``b",
``c", and ``d" states according to their peak luminosity of synchrotron radiation.

\emph{Mrk 501.} It is a TeV source with $z=0.034$, which was identified by Whipple during an observation
campaign of 66 hr. An average flux of $(8.1\pm1.4)\times10^{-12}$cm$^{-2}$s$^{-1}$ above 300 GeV and flux
variation on timescale of days was observed (Quinn et al. 1996). From 1997 April to 1999 June, this source
was detected quasi-simultaneously with BeppoSAX and the Cerenko Array. The observations show that the peak
frequency of the synchrotron emission shifts from 100 keV back to 0.5 keV, and correspondingly the flux
decreased (Tavecchio et al. 2001). The SEDs ``a" and ``b" of this source are taken from the broadband
observations in 1997 April 16th and 1997 April 7th.  The data of the SED ``c" is taken from a 4.5 month-long multi-wavelength observation campaign for
Mrk 501 performed from 2009 March 15 to August 1 (Abdo et al. 2011)\footnote{Note that the TeV data of this SED has been corrected for the extra-galactic light (EBL) absorption with a model proposed by Franceschini et al. (2008), so we do not make further EBL correction in our SED fit.}. The SED marked as ``d" is taken from a multi-wavelength campaign with Suzaku and MAGIC performed in 2006 July. The average VHE flux
above 200 GeV in this campaign  is $\sim 20\%$ of the Crab flux, with a photon index $2.8\pm0.1$ from 80 GeV to 2 TeV (Anderhub et
al. 2009). The data in the GeV band for this SED are taken from {\em Fermi}/LAT observations (Abdo et al. 2009).

\emph{3C~279.} The redshift of this sources is $0.536$. It is the first bright $\gamma$-ray blazar reported by
the EGRET instrument aboard the Compton Gamma-Ray Observatory (Wehrle et al. 1998). It is also the first
confirmed TeV FSRQ (Albert et al. 2008). Its SED ``a" is obtained from a monitoring program with the Compton Gamma Ray observatory, the Ginga
X-ray satellite, and instruments in other bands performed in 1991 June during a $\gamma$-ray flare (Hartman
et al. 1996). The SED ``b" is from observations in 2006 February (Berger et al. 2011). The SED ``c" and ``d" are obtained during this source experienced a sharp gamma-ray flare and an isolated X-ray flare reported by Abdo et
al. (2010), respectively.

\section{Model and SED Fits}
As shown in Zhang et al. (2012a), the external photon fields from the BLRs can be neglected in comparison
with the synchrotron radiation photon fields for BL Lacs. Therefore, the syn+SSC model is used to fit the SEDs of Mrk 501.
For FSRQ 3C~279, the contribution of external photon field from the BLR cannot be neglected, and thus the syn+SSC+EC model is used to fit the SEDs of 3C~279. The total luminosity of the BLR for 3C~279 is taken from Celotti et al. (1997) and the size of the BLR is calculated
using the BLR luminosity with equation (23) presented in Liu \& Bai (2006). The radiation from a BLR is taken as
a black body spectrum and the corresponding energy density measured in the co-moving frame is $U^{'}_{\rm
BLR}=5.37\times10^{-3}\Gamma^{2}$ erg cm$^{-3}$, where the Lorentz factor of the jet is taken as its
Doppler boosting factor, i.e., $\Gamma=\delta$, because the relativistic jets of blazars are close to the line
of sight. The minimum variability timescales are taken as $\Delta t=1$ hr for Mrk 501 (Tavecchio et al. 2001)
and $\Delta t=8$ hr for 3C~279 (Wehrle et al. 1998). The energy distributions of the relativistic electrons in
the radiating regions are assumed to be a broken power-law with indices $p_1$ and $p_2$ as well as a break at
$\gamma_{\rm b}$ within the energy range [$\gamma_{\rm min}$, $\gamma_{\rm max}$]. The details of the models and the
strategy for parameter constraints are presented in Zhang et al. (2012a).

The SEDs of Mrk 501 and 3C~279 are well fitted with the models. Our best SED fits and the contours of the probability $p$ ($p\propto
e^{-\chi^2/2}$) for the parameters $B$ and $\delta$ at 1 $\sigma$ significance level are also presented in
Figure \ref{SED}. The contours are thin and slanted ellipses, especially for
Mrk 501, which are consistent with that reported by Zhang et al. (2012a). The EC components of SEDs for 3C~279
present further constraint on $\delta$, hence make tighter constraint on $B$ and $\delta$ than that for Mrk~501.
The parameters of the best fit to the SED as shown in Figure 1 are marked with stars in the contours. We also calculate the bolometric luminosities ($L_{\rm bol}$) with our best SED fits. Our results are reported in Table
1. One can observe that $\gamma_{\rm b}$ ranges in $10^4-10^6$ for Mrk~501, but it
is $200\sim 600$ for 3C~279. The values of $\delta$ for the two sources are
comparable within errors. The magnetic field strength of Mrk~501 is roughly one order of magnitude lower than
that of 3C~279.

\section{Observed Spectral Variation and the Possible Physical Origins}
\subsection{Spectral Variation of Synchrotron Radiation Bump}
\begin{figure}
\includegraphics[angle=0,scale=0.2]{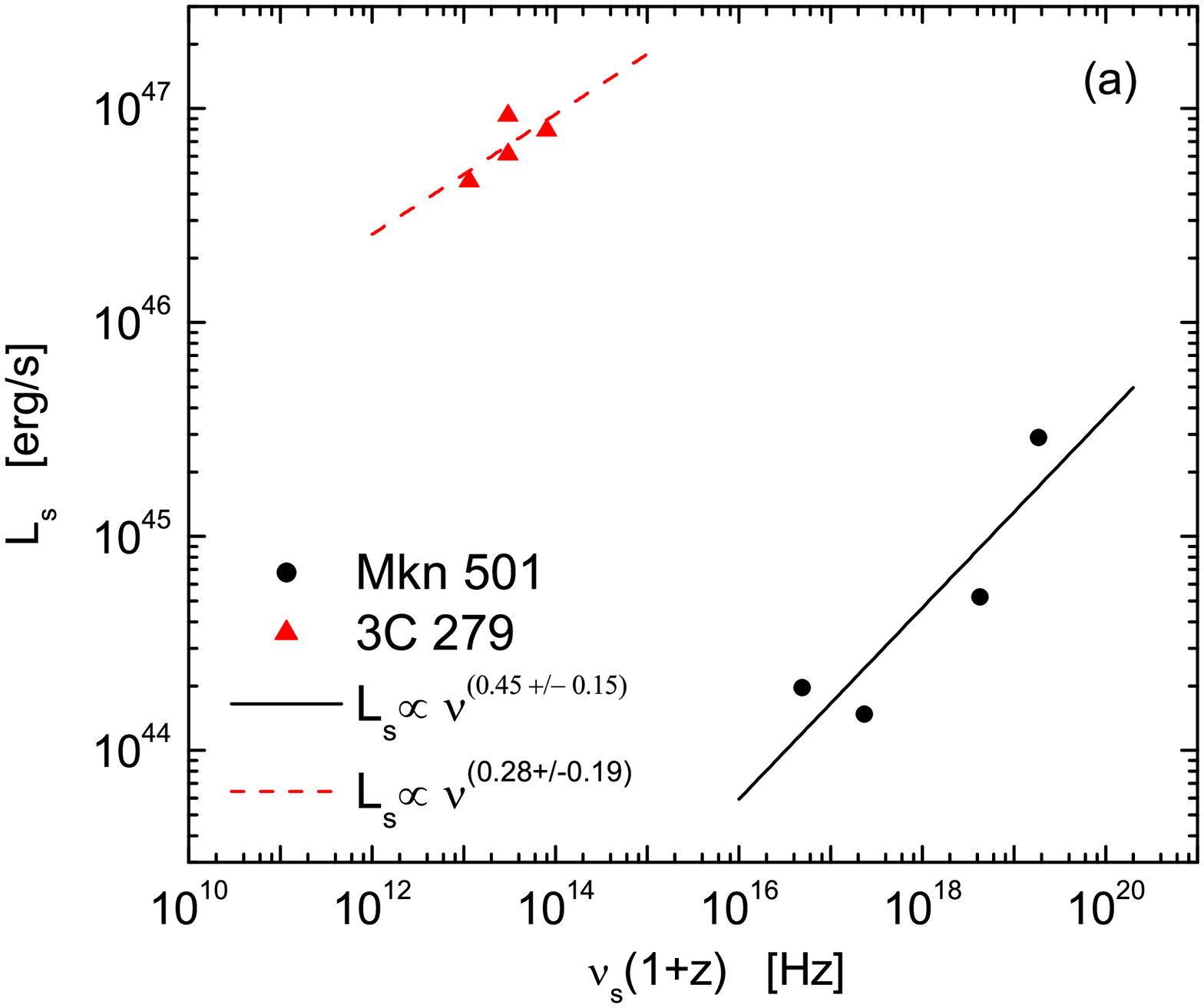}
\includegraphics[angle=0,scale=0.2]{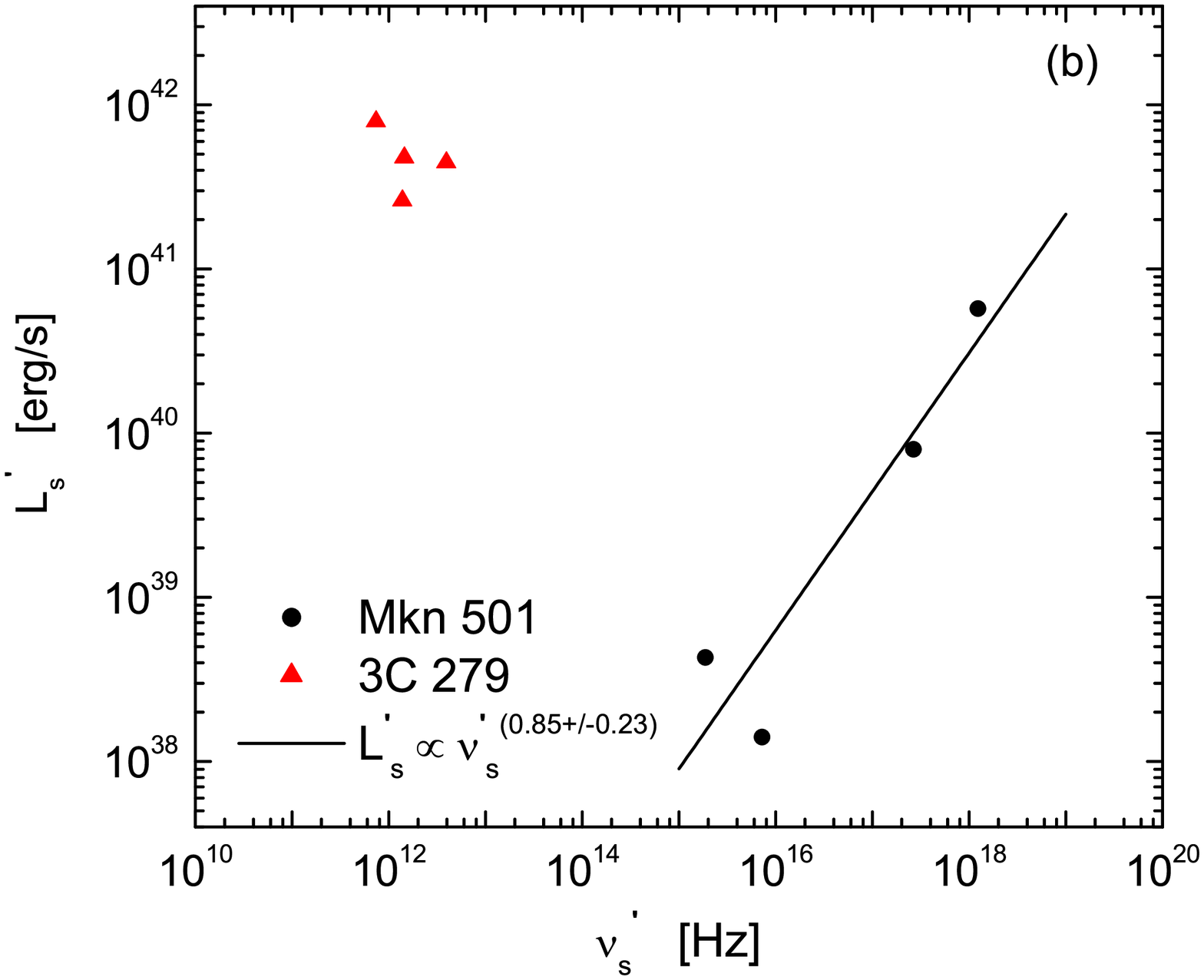}
\caption{Peak frequency of synchrotron radiation as a function of the peak luminosity in the observer ({\em Panel a}) and co-moving ({\em Panel b}) frames. The best fit lines and the slopes are also shown in each panel.}\label{Ls_nus}
\end{figure}

\begin{figure}
\includegraphics[angle=0,scale=0.35]{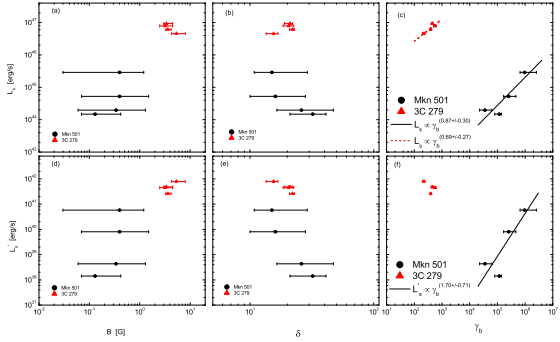}
\caption{Magnetic filed strength $B$, the beaming factor $\delta$, and the break Lorentz factor of electron
$\gamma_{\rm b}$ as a function of the peak luminosity of synchrotron radiation in both the observer ($L_{\rm s}$)
and co-moving ($L^{'}_{\rm s}$) frames for Mrk 501 and 3C~279.}\label{Ls_Para}
\end{figure}

The observed peak luminosity ($L_{\rm s}$) of the synchrotron radiation as a function of the corresponding peak frequency ($\nu_{\rm s}$) for the SEDs in our sample is shown in Figure \ref{Ls_nus} (a). By making the beaming factor correction\footnote{Here we do not consider the
errors of $\delta$ and just use the central value of $\delta$ to correct the parameters.} for $\nu_{\rm
s}$ and $L_{\rm s}$, we show the relation of $\nu_{\rm s}-L_{\rm s}$ in the co-moving frame in Figure \ref{Ls_nus} (b), where the prime denotes the data in the co-moving frame.  To study the dependence of $L_{\rm s}$ and $L^{'}_{\rm s}$ on the physical parameters of the jets, we show $L_{\rm s}$ (or $L^{'}_{\rm s}$) as functions of $B$, $\delta$, and $\gamma_{\rm b}$ in Figure \ref{Ls_Para}. The results of pair correlation analysis with the Pearson method are reported in Table 2\footnote{Considering the small sample statistics and large errors of the parameters, we also use a bootstrap method to examine the correlation coefficients of these correlations. The results are consistent with that of the Pearson correlation analysis and are also shown in Table 2.}.

\begin{figure}
\includegraphics[angle=0,scale=0.35]{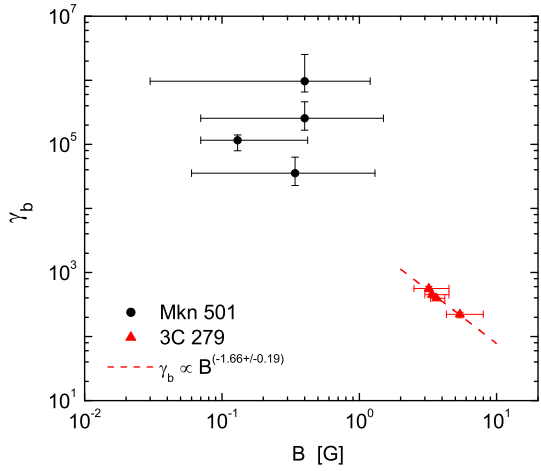}
\caption{Break Lorentz factor of electron $\gamma_{\rm b}$ as a function of the magnetic filed strength $B$
for Mrk 501 and 3C~279.}\label{B_gammab}
\end{figure}

\begin{itemize}
\item Mrk 501. A clear correlation between  $L_{\rm s}$ and $\nu_{\rm
s}$ in the observer frame is found, i.e., $L_{\rm s}\propto\nu_{\rm s}^{0.45\pm0.15}$, which is consistent with that observed in Mrk 421 ($L_{\rm s}\propto\nu_{\rm s}^{0.42\pm0.06}$, Tramacere et al. 2009). $L^{'}_{\rm s}$ is also correlated with $\nu^{'}_{\rm s}$, but the slope changes to $0.85\pm0.23$. No clear dependence of $L_{\rm s}$ and $L^{'}_{\rm s}$ on $B$ and $\delta$ is observed. However, both $L_{\rm s}$ and $L^{'}_{\rm s}$ are correlated with $\gamma_{\rm b}$, i.e., $L_{\rm s}\propto \gamma_{\rm b}^{0.87\pm0.30}$ and $L^{'}_{\rm s}\propto \gamma_{\rm b}^{1.70\pm0.71}$. These results suggest that the observed $\nu_{\rm s}-L_{\rm s}$ correlation could be dominated by the variation of $\gamma_{\rm b}$.

\item 3C 279. The variations of $L_{\rm s}$ and $L^{'}_{\rm s}$ of 3C 279 are much smaller than that of Mrk 501. Although a tentative correlation between $L_{\rm s}$ and  $\nu_{\rm s}$ is also observed, the slope of the correlation is much shallower than that in Mrk 501, i.e., $L_{\rm s}\propto\nu_{\rm s}^{0.28\pm0.19}$. Even more, we do not find any correlation between $L^{'}_{\rm s}$ and $\nu^{'}_{\rm s}$. This is reasonable since $\gamma_{\rm b}$ of 3C 279 in different states is in a very narrow range. We do also not find any dependence of $L^{'}_{\rm s}$ on $\gamma_{\rm b}$, $B$, and $\delta$ as shown in Figure \ref{Ls_Para}. Tentative $\gamma_{\rm b}-L_{\rm s}$ correlation and $B-L_{\rm s}$ anti-correlation are found in 3C 279. We infer that they may be from an anti-correlation between $B$ and $\gamma_{\rm b}$, as shown in Figure \ref{B_gammab}. A clear anti-correlation of $B-\gamma_{\rm b}$ is observed in 3C 279, which is $\gamma_{\rm b}\propto B^{-1.66\pm 0.19}$. The observed anti-correlation of $B-\gamma_{\rm b}$ in 3C 279 is consistent with the radiation cooling for the electrons by the magnetic field.
\end{itemize}
Similar $\nu_{\rm s}-L_{\rm s}$ correlation was also found by Massaro et al.(2008) for some high-frequency-peaked BL Lacs (see also Zhang et al. 2012a). They proposed that this correlation is a signature of synchrotron radiation. Under the mono-frequency approximation condition, we get $L_{\rm s}\propto N\gamma_{\rm b}^2B^2\delta^4$ and $\nu_{\rm s}\propto \gamma_{\rm b}^2B\delta$, where $N$ is the number of emitting electrons. The comparison between the observed correlations and the model predictions can provide some information about which parameter may dominate the observed correlations. One can infer that $L^{'}_{\rm s}\propto \nu^{'}_{\rm s}$ if $\gamma_{\rm b}$ dominates the relation of $\nu^{'}_{\rm s}-L^{'}_{\rm s}$ and $L^{'}_{\rm s}\propto {\nu^{'2}_{\rm s}}$ if $B$ dominates the relation of $\nu^{'}_{\rm s}-L^{'}_{\rm s}$ (see also Massaro et al. 2008). We find that $L^{'}_{\rm s}\propto {\nu_{\rm s}^{'0.85\pm0.23}}$ for Mrk 501, indicating that the relation of $\nu^{'}_{\rm s}-L^{'}_{\rm s}$ in Mrk 501 should be due to the variation of $\gamma_{\rm b}$. However, no significant spectral variation of synchrotron process bump is observed for 3C 279, especially in the co-moving frame.

\subsection{Spectral Variation of the IC Bump}
\begin{figure}
\includegraphics[angle=0,scale=0.2]{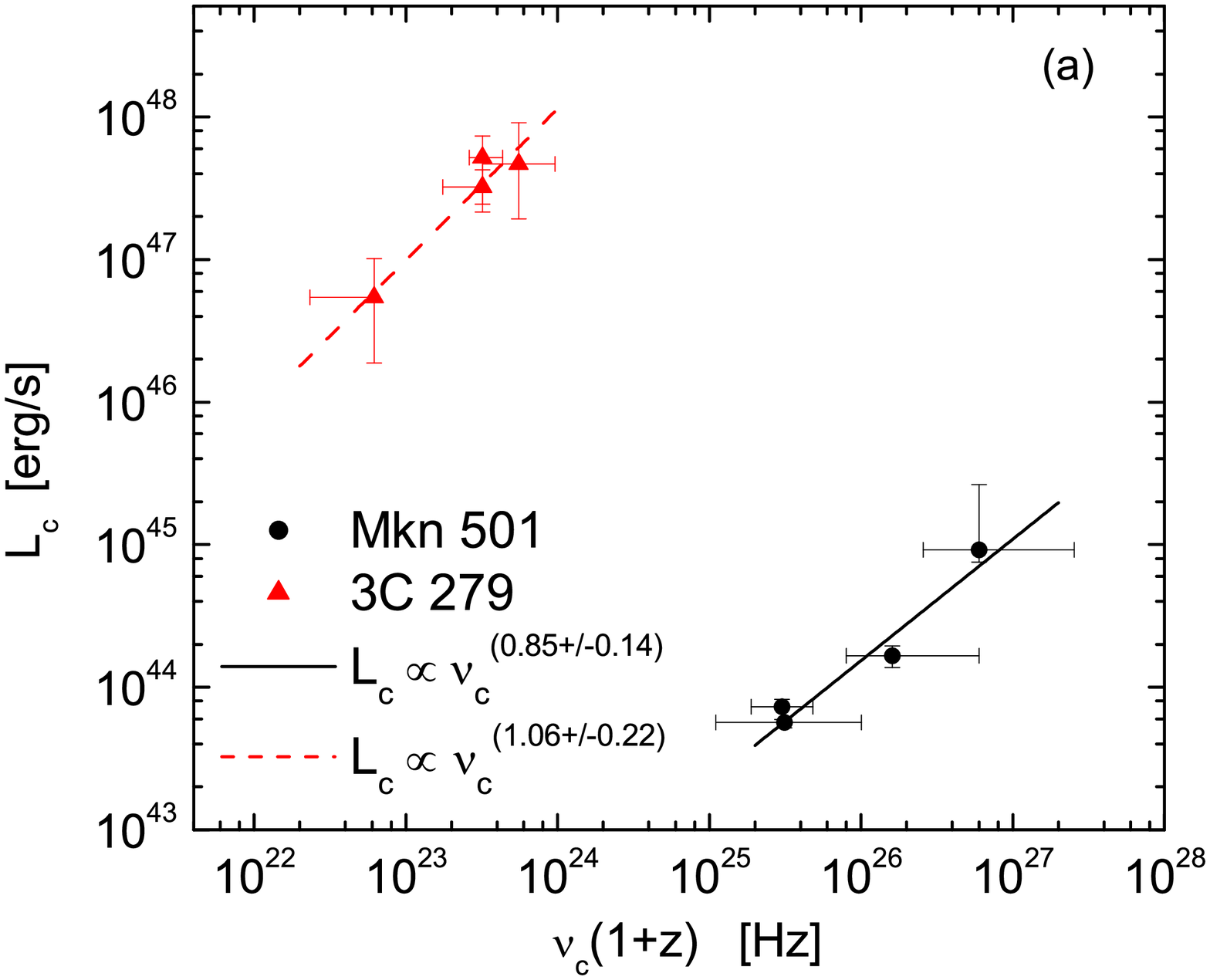}
\includegraphics[angle=0,scale=0.2]{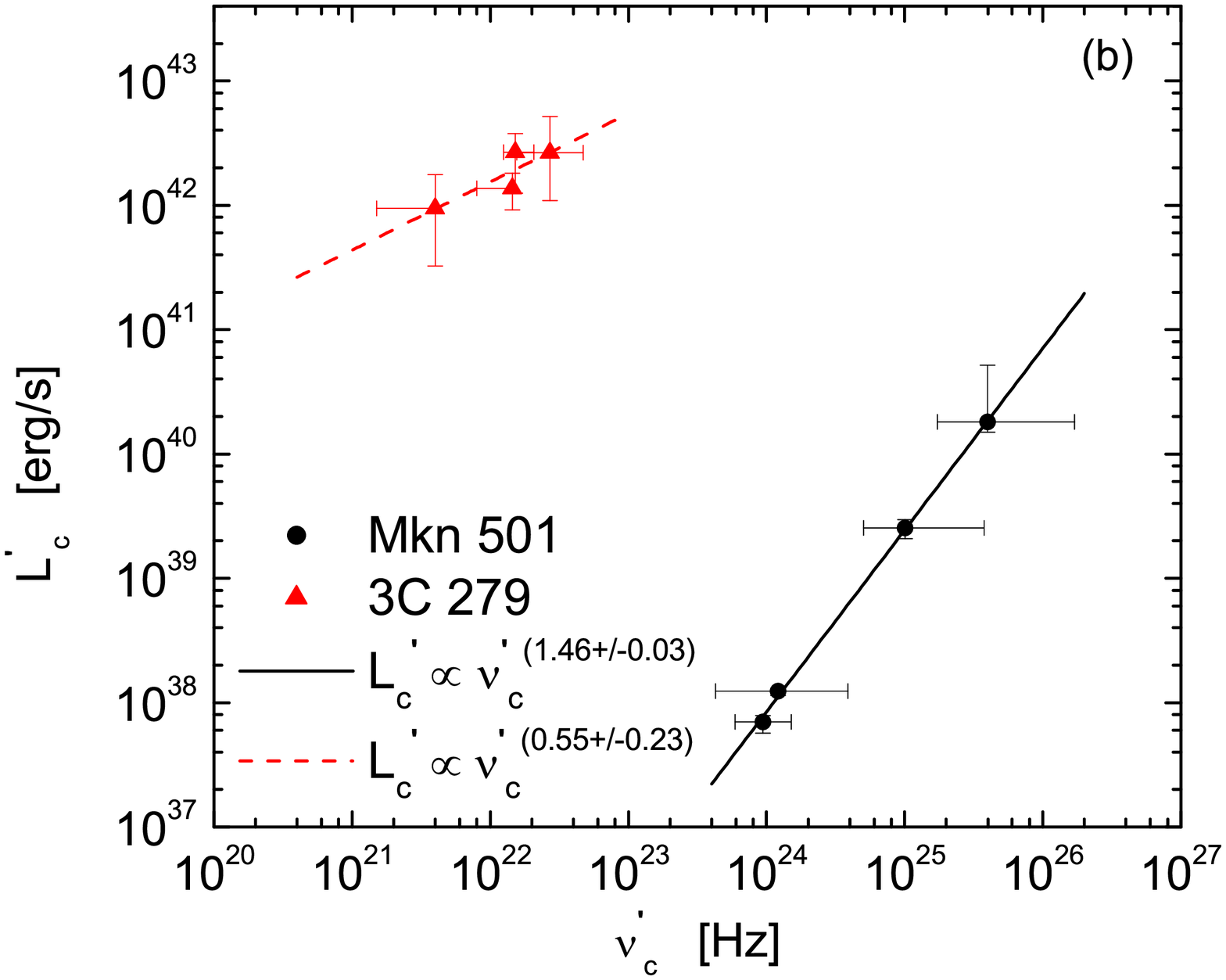}
\caption{Peak frequency of IC process as a function of the peak luminosity in the observer
({\em Panel a}) and in the co-moving ({\em Panel b}) frames.}\label{Lc_nuc}
\end{figure}

\begin{figure}
\includegraphics[angle=0,scale=0.35]{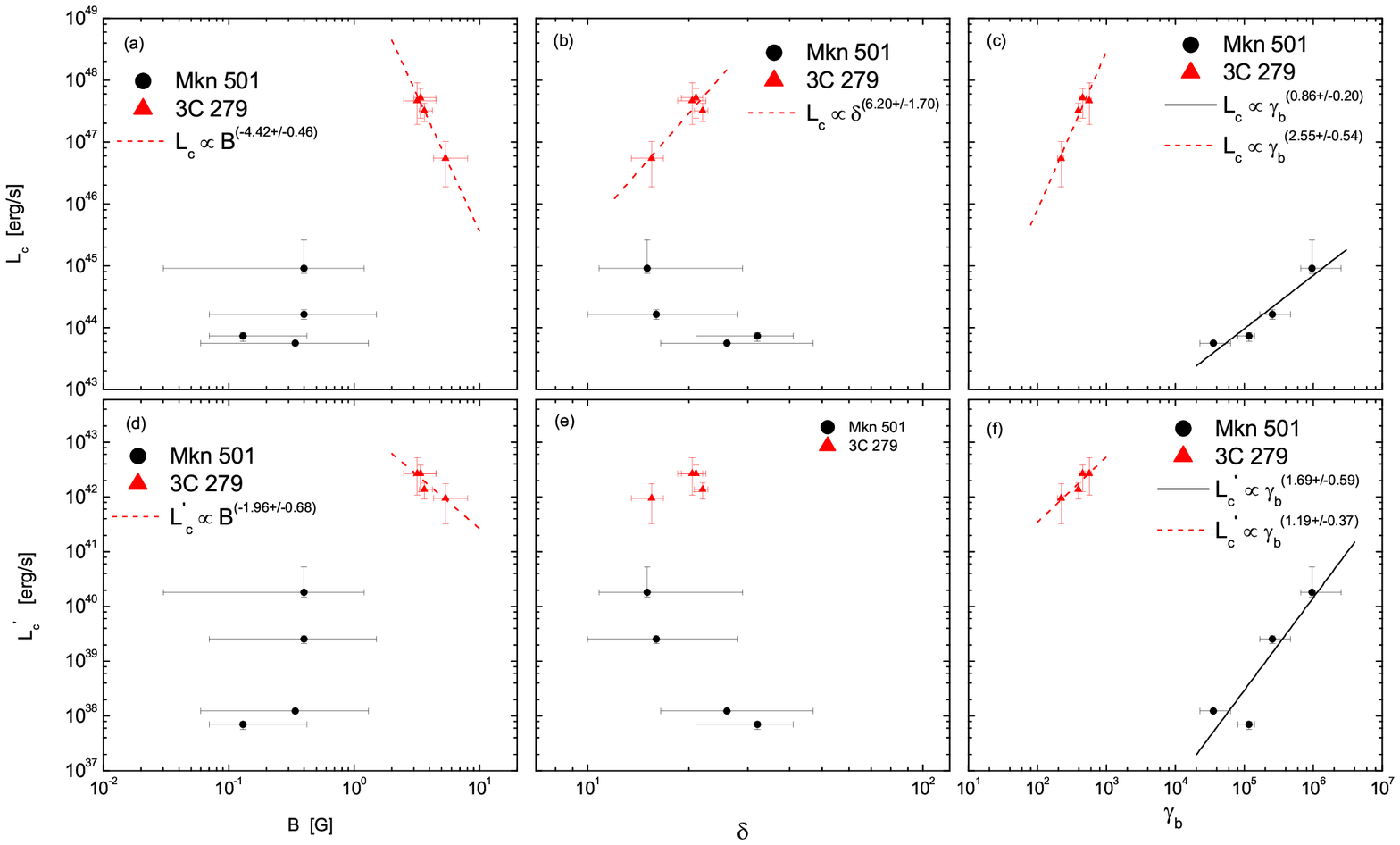}
\caption{Magnetic filed strength $B$, the beaming factor $\delta$, and the break Lorentz factor of electron
$\gamma_{\rm b}$ as a function of the peak luminosity of the IC bump in both the observer ($L_{\rm c}$) and co-moving ($L^{'}_{\rm c}$)
frames for Mrk 501 and 3C~279.}\label{Lc_Para}
\end{figure}

We present similar analysis for the IC bump as that for the synchrotron radiation bump. The IC bump is dominated by the SSC process for Mrk 501 and the EC process for 3C 279, therefore, we give the analysis of the SSC and EC bumps for Mrk 501 and 3C 279, respectively. Figure \ref{Lc_nuc} shows the correlations of $\nu_{\rm c}-L_{\rm c}$ in the observed and co-moving frames and Figure \ref{Lc_Para} displays $L_{\rm c}$ (or $L^{'}_{\rm c}$) as functions of $B$, $\delta$, and $\gamma_{\rm b}$ for the two sources. The Pearson correlation analysis results are also reported in Table 2.

\begin{itemize}
\item Mrk 501. It is found that $L_{\rm c}$ is correlated with $\nu_{\rm c}$ in both the observer and co-moving frames. The best fits give $L_{\rm c} \propto \nu_{\rm c}^{0.85\pm 0.14}$ and $L^{'}_{\rm c} \propto \nu_{\rm c}^{'1.46\pm 0.03}$. The power-law index of the observed $\nu_{\rm c}-L_{\rm c}$ relation is much larger than that of the $\nu_{\rm s}-L_{\rm s}$ relation, especially in the co-moving frame. This is inconsistent with the SSC model. Note that the IC peak at the TeV gamma-ray band would be affected by the Klein-Nishina (KN) effect, which would result in a larger slope in the observed $\nu_{\rm c}-L_{\rm c}$ relation than that in the $\nu_{\rm s}-L_{\rm s}$ relation, especially for a transition from the Thomson regime to the KN regime (Tramacere et al. 2011). The IC peak of SED ``a" for Mrk 501 is in the Thomson regime and the other three SEDs are in the KN regime. Therefore, the discrepancy of the slopes in the IC process and the synchrotron radiation bump may be due to the KN effect. Both $L_{\rm c}$ and $L_{\rm c}^{'}$ are correlated with $\gamma_{\rm b}$. The best fits give $L_{\rm c}\propto \gamma_{\rm b}^{0.86\pm0.20}$ and $L^{'}_{\rm c}\propto \gamma_{\rm b}^{1.69\pm0.59}$. No clear dependence of $L_{\rm c}$ and $L^{'}_{\rm c}$ on $B$ and $\delta$ is found. Therefore, the variation of the IC peak for this source should be also due to the variation of $\gamma_{\rm b}$ as that for the synchrotron peak.

\item 3C 279. Its IC peak is dominated by the EC process. Its $\nu_{\rm c}$ is also correlated with $L_{\rm c}$ in both the observer and co-moving frames. The best fits give $L_{\rm c} \propto \nu_{\rm c}^{1.06\pm0.22}$ and $L^{'}_{\rm c} \propto \nu_{\rm c}^{'0.55\pm0.23}$. Note that the slopes of relation $\nu_{\rm c}-L_{\rm c}$ in 3C 279 and Mrk 501 are roughly consistent within the error bars, but the slope of $\nu^{'}_{\rm c}-L^{'}_{\rm c}$ in 3C 279 is much smaller than that in Mrk 501. This is because the IC bump of 3C 279 is dominated by EC process and the select SEDs of 3C 279 at gamma-ray band are not affected by the KN effect. $L_{\rm c}$ is strongly dependent on $\delta$, i.e., ${L_{\rm c}}\propto \delta^{6.20\pm 1.70}$, but $L^{'}_{\rm c}$ is not, indicating that the Doppler boosting factor is a great ingredient for the variation of $L_{\rm c}$. Both $L_{\rm c}$ and $L^{'}_{\rm c}$ are also anti-corrected with $B$, i.e., ${L_{\rm c}}\propto B^{-4.42\pm 0.46}$ and ${L^{'}_{\rm c}}\propto B^{-1.96\pm 0.68}$. This implies that the magnetic field may also play an important role in the spectral variation of 3C 279. Similar to Mrk 501, both $L_{\rm c}$ and $L^{'}_{\rm c}$ are corrected with $\gamma_{b}$ for 3C 279, i.e., ${L_{\rm c}}\propto \gamma_{\rm b}^{2.55\pm 0.54}$ and ${L^{'}_{\rm c}}\propto \gamma_{\rm b}^{1.19\pm0.37}$. This suggests that $\gamma_{\rm b}$ is also a key factor for the variation of the EC peak, and is anticorrelated with $B$, being consistent with the radiation cooling for the electrons by the magnetic field.
\end{itemize}
As shown above, the variation of the IC peak for Mrk 501 is also dominated by the electron spectrum, similar to its synchrotron peak. The variation of the IC peak for 3C 279 depends on $\delta$, $B$, and $\gamma_{\rm b}$. The electron spectrum is critical since both the synchrotron radiation and the IC process are highly dependent on it. Significant variabilities in both the synchrotron process and IC
bump of Mrk 501 are related to the variation of $\gamma_{\rm b}$, which ranges within $10^{4}\sim 10^{6}$, implying that the
particle acceleration in different epoches of Mrk 501 would be different. Note that the environments of the emitting regions of BL Lacs and FSRQs are dramatically different. The radiating region of 3C~279 is suggested to be inside the BLR. The electrons thus would be cooled down by the EC process. On the other hand, the magnetic field strength in 3C 279 is usually one order magnitude higher than that in Mrk 501 and the electrons may also be cooled through the radiation cooling effect by the magnetic field. These effects would result in a narrow distribution of $\gamma_{\rm b}$ (i.e., $\gamma_{\rm b}=200\sim 600$) among different states for 3C 279, and thus there is no significant variation in the synchrotron bump. The variation of the IC peak for 3C~279 is dominated by the EC process. The energy density of the external photon field is magnified by $\Gamma^2$ ($\sim\delta^2$) and the energy of seed photon is magnified by $\Gamma$ ($\sim\delta$) in the co-moving frame, thus a small variation of $\delta$ would result in significant variations of $L^{'}_{\rm c}$ and $\nu^{'}_{\rm c}$. Hence, the variation of IC peak for 3C 279 is dominated by the variation of $\delta$. However, the variation of $B$ may be also responsible for the variation of IC peak, which result in a correlation of $\gamma_{\rm b}-L_{\rm c}$ (or $\gamma_{\rm b}-L^{'}_{\rm c}$) because of the cooling effect.

 \section{Discussion}
It was found that high-luminosity FSRQs tend to have a low peak frequency and low-luminosity BL Lacs tend to have a high peak frequency. This is the so-called blazar sequence, which may be related to the different environments of emitting regions for different types of blazars (e.g., Ghisellini et al. 1998). With the discovery of ``blue quasars", Ghisellini \& Tavecchio (2008) further proposed that, more physically, the blazar sequence is due to the
difference of the black hole (BH) masses and accretion rates among these sources. In this scenario, the ``blue quasars" should have a large BH mass and an intermediate accretion rate. Their emission regions are beyond the BLR. Moreover, the red low-luminosity blazars should exist, which may have a small BH mass and a relatively large accretion rate. Chen \& Bai (2011) reported that the properties of the narrow line Seyfert 1 galaxies, which have a low-peak-frequency and low-luminosity, are similar to this kind of blazar. As presented in Figures \ref{Ls_nus} and \ref{Lc_nuc}, the data points of 3C 279 are in the left-top and the data points of Mrk 501 are in the right-bottom of the figures, illustrating the feature of the blazar sequence. The observed $\nu_{\rm s}-L_{\rm s}$ correlation in individual blazars is opposite to the blazar sequence and is referred as {\em blazar anti-sequence} of spectral variation in individual blazars (Zhang et al. 2012b).

The results in this analysis indicate that the environment effect may be also fundamental for the observed different spectral variations
of 3C~279 and Mrk~501. The significant differences of the electron energy distributions between Mrk 501 and 3C 279 may be mainly due to the cooling effect of the external photons from the BLR of 3C 279. However, the spectral variation among different states of an individual source is independent of its environment. As shown in our analysis the variation of the electron spectrum may result in the spectral variation of BL Lacs, and the bulk motion of the emitting region as well as the magnetic field strength would be responsible for the spectral variation of FSRQs.

Different from the blazar sequence, the rapid spectral variation in individual blazars would be not originated from the variation of the accretion rate. The variability timescale produced by the variations of the accretion rate can be estimated by the viscous timescale of disk,
\begin{equation}
t_{\rm visc}\sim 3\times10^5 \alpha^{-4/5}\dot{M}_{16}^{-3/10}m^{1/4}_{1}R_{10}^{5/4}  \;\;\;\rm s,
\end{equation}
where $R_{10}=R/(10^{10}$ cm), $\dot{M}_{16}=\dot{M}/(10^{16}$ g s$^{-1})$, $m_1=M/M_{\bigodot}$. Taking the
typical values of those parameters for blazars, $M_{\rm BH}=10^{8.5}M_{\bigodot}$, $\dot{M}=0.1\dot{M}_{\rm
Edd}$, $R=10R_{\rm s}$, and $\alpha=0.1$, we get $t_{\rm visc}=6.03\times10^{6}$ days, which is much larger than
the observed variability timescales of blazars. The instability of the accretion disk cannot produce the rapid variability as
that observed in blazars. On the other hand, the variation of the accretion rate may result in the luminosity variations of both the disk and the broad emission lines. We cannot test this scenario with Mrk 501 since BL Lacs have no or very weak emission lines and their disk luminosity is also very weak, which may be much lower than the non-thermal radiation from the jet. It is found that the strength of Ly$\alpha$ of 3C~279 is nearly constant when the continuum flux varies, as that also observed in 3C 273 (Koratkar et al. 1998). Therefore, the blazar anti-sequence of spectral variation in individual blazars would be not originated from the change of the accretion rate.

The rapid variability of blazars should be dominated by the jet emission. It is generally believed that jet launching is connected with the accretion disk and corona
in a source (Armitage \& Natarajan 1999; Merloni \& Fabian 2002; Cao 2004, Chen et al. 2012). Using the observations of
BeppoSAX, Grandi \& Palumbo (2004) untangled the emissions of jet, accretion disk, and corona for 3C 273, and
found that the emission of corona becomes weak when the jet emission increases in X-ray band, implying that the variation of the jet emission is related with the variation of the corona emission. Therefore, the instability of the corona may result in the variation of the physical condition of jets and lead to the variation of the jet emission. On the other hand, in a BH X-ray binary, the X-ray spectrum becomes soft when the corona-dominated emission drops, and the properties of its jet, such as density and magnetic field, is also changed (Fender \& Belloni 2012). This is consistent with our results, implying that the jet emission of
blazars may be also related with the corona emission. The dominated mechanism of corona variation may be magnetic reconnections, like the corona activity in the Sun (Zhang et al. 2000; Zhang 2007). For a super-massive black hole, the timescale of magnetic reconnections in the disk is from hours to days, which is consistent with the
observed variability timescale of blazars.

\section{Summary}

SEDs observed at four epochs for Mrk 501 and 3C 279 are compiled from literature to investigate the spectral variation in individual blazars. Our results are summarized below.

\begin{itemize}
\item The SEDs of Mrk 501 and 3C 279 are well fitted with the syn+SSC and syn+SSC+EC models, respectively. The derived values of $\gamma_b$ are $10^4-10^6$ for Mrk 501 and $200\sim 600$ for 3C 279. The magnetic field strength of Mrk 501 is usually one order of magnitude lower than that of 3C 279. The Doppler factors of the two sources are comparable.

\item For Mkn 501, the peak luminosities of the synchrotron radiation and IC bumps are tentatively correlated with the corresponding peak frequencies and they are also correlated with $\gamma_b$ in both the observer and co-moving frames, but they are not correlated with $B$ and $\delta$.

\item For 3C 279, no significant variation is observed for the synchrotron radiation peak and its luminosity variation is dominated by the EC peak. Its EC peak luminosities in both the observer and co-moving frames are correlated with the peak frequencies, are also correlated with $\gamma_{\rm b}$ and $\delta$, but anti-correlated with $B$.
\end{itemize}

These results suggest that the variation of the electron spectrum may result in the spectral variation of Mrk 501, and more complicated, the variations of the electron spectrum, the bulk motion of emitting region, as well as the magnetic field strength would be responsible for the spectral variation of 3C 279. However, the cooling effect by external field photons may smear out the difference of electron spectrum in different states of 3C 279 and its spectral variation would be dominated by the EC component, hence the magnification of the external photon field by the bulk motion of the jet is an essential reason for the spectral variation.

Our results show that the spectral variation in individual blazars is different from the blazar sequence for different types of blazars. We refer this spectral variation feature as blazar anti-sequence of spectral variation in individual blazars. Based on our brief discussion, we propose that this spectral variation feature may be originated from the instability of the corona but not from the variation of the accretion rate as that for the blazar sequence.

\acknowledgments This work is supported by the National Basic Research Program (973 Programme) of China (Grant
2009CB824800), the National Natural Science Foundation of China (Grants 11078008, 11025313, 11133002, 10725313),
China Postdoctoral Science Foundation, Guangxi Science Foundation (2011GXNSFB018063, 2010GXNSFC013011), Guangxi
SHI-BAI-QIAN project (Grant 2007201), and the Qianren start-up grant 292012312D1117210.

\clearpage
\begin{deluxetable}{llllllllllll}
\tabletypesize{\footnotesize}\tablecolumns{11}\tablewidth{36pc} \tablecaption{Observations and SED fit results
for Mrk 501 and 3C~279}\tablenum{1} \tablehead{\colhead{Source} & \colhead{State\tablenotemark{a}} &
\colhead{$p_{1}$}
& \colhead{$p_{2}$}  & \colhead{$\gamma_{\rm min}$} & \colhead{$\gamma_{\rm b}$} & \colhead{$\gamma_{\rm max}$}& \colhead{$N_{0}$}&\colhead{$\delta$} & \colhead{$B$} & \colhead{$\delta_{\tau}\tablenotemark{b}$} & \colhead{$L_{\rm bol}$} \\
\colhead{} & \colhead{} & \colhead{}& \colhead{} & \colhead{} & \colhead{} & \colhead{} & \colhead{} &
\colhead{} &
\colhead{(G)} & \colhead{}&  \colhead{(erg/s)} \\
\colhead{(1)}& \colhead{(2)} & \colhead{(3)} & \colhead{(4)}& \colhead{(5)} & \colhead{(6)} & \colhead{(7)} &
\colhead{(8)} &\colhead{(9)} &\colhead{ (10)} & \colhead{(11)}& \colhead{(12)}} \startdata
Mrk 501&a&1.86&4.6&2&9.65E5&1.5E8&1E4&15$^{29}_{10.8}$&0.4$^{1.2}_{0.03}$&10.8&2E46\\
 &b&2&3.2&20&2.56E5&9E6&2.2E4&16$^{28}_{10}$&0.4$^{1.5}_{0.07}$&9.9&5.5E45\\
  &c&2.24&3.72&300&3.56E4&1E6&3.9E4&26$^{47}_{16.5}$&0.34$^{1.3}_{0.06}$&11.6&1.7E45\\
 &d&2.64&3.3&900&1.16E5&8E5&1.7E6&32$^{41}_{21}$&0.13$^{0.42}_{0.07}$&11.1&1.9E45\\
3C~279&a&2.2&4.2&2&451&3E3&1.3E5&21$^{22}_{19}$&3.4$^{4.5}_{3}$&5.5&4.7E48\\
 &b&2.2&4.2&2&563&7E4&1.2E5&20.5$^{22.5}_{18.5}$&3.2$^{4.5}_{2.5}$&13.1&3.7E48\\
&c&2.2&4.46&2&394&3E3&5.8E4&22$^{22.8}_{20.8}$&3.65$^{4.2}_{3.3}$&4.6&2.3E48\\
&d&2.1&4.3&45&223&1E4&2.1E5&15.5$^{16.8}_{13.5}$&5.4$^{8}_{4.3}$&4.3&5.6E47\\
\enddata
\tablenotetext{a}{The state of each SED, which is defined according to the peak luminosity of the synchrotron
radiation bump.} \tablenotetext{b}{It is calculated using equation (23) given in Tavecchio et al. (1998).}
\tablecomments{Columns: (3) (4) The energy indices of electrons below and above the break; (5) (6) (7) The
minimum, break, and maximum Lorentz factors of electrons; (8) The electron density parameter $N_0$; (9) The
beaming factor $\delta$; (10) The magnetic field strength $B$; (11) The lower limit of $\delta$ derived by the
$\gamma$-ray transparency condition; (12) The bolometric luminosity of each SED. }
\end{deluxetable}

\begin{deluxetable}{lcccccccccc}
\tabletypesize{\scriptsize}\tablecolumns{9}\tablewidth{40pc} \tablecaption{Results of the best linear fits with
the Pearson correlation analysis method for the parameter sets ($x-y$) of Mrk~501 and 3C~279. The fitting
function is $\log y=a+b\log x$. The Spearman correlation coefficient is $r$ and the chance probability is
$p$.}\tablenum{2} \tablehead{\colhead{} &
\multicolumn{5}{c}{Mrk 501}&\multicolumn{5}{c}{3C~279}\\\cline{2-6}\cline{7-11}\\
\colhead{}&\colhead{$a$}&\colhead{$b$}&\colhead{$r$}&\colhead{$p$}&\colhead{$\tablenotemark{1}r_{\rm
bootstrap}$}&\colhead{$a$}&\colhead{$b$}&\colhead{$r$}&\colhead{$p$}&\colhead{$\tablenotemark{1}r_{\rm
bootstrap}$}} \startdata
$\nu_{\rm s}-L_{\rm s}$&36.6$\pm$2.8&0.45$\pm$0.15&0.90&0.10&\nodata&43.0$\pm$2.6&0.28$\pm$0.19&0.72&0.23&\nodata\\
$\nu^{'}_{\rm s}-L^{'}_{\rm s}$&25.3$\pm$3.8&0.85$\pm$0.23&0.94&0.06&\nodata&\nodata&\nodata&\nodata&\nodata&\nodata\\
$\gamma_{\rm b}-L_{\rm s}$&40.1$\pm$1.6&0.87$\pm$0.30&0.90&0.10&\nodata&45.1$\pm$0.7&0.69$\pm$0.27&0.87&0.13&\nodata\\
$\gamma_{\rm b}-L^{'}_{\rm s}$&30.4$\pm$3.8&1.70$\pm$0.71&0.86&0.14&\nodata&\nodata&\nodata&\nodata&\nodata&\nodata\\
$\nu_{\rm c}-L_{\rm c}$&22.1$\pm$3.6&0.85$\pm$0.14&0.97&0.03&0.90&22.5$\pm$5.2&1.06$\pm$0.22&0.96&0.04&0.96\\
$\nu^{'}_{\rm c}-L^{'}_{\rm c}$&2.8$\pm$0.9&1.46$\pm$0.03&$\sim1.0$&5.6E-4&0.96&30.1$\pm$5.1&0.55$\pm$0.23&0.86&0.14&0.83\\
$B-L_{\rm c}$&\nodata&\nodata&\nodata&\nodata&\nodata&50.0$\pm$0.3&-4.42$\pm$0.46&-0.99&0.01&-0.87\\
$B-L^{'}_{\rm c}$&\nodata&\nodata&\nodata&\nodata&\nodata&43.4$\pm$0.4&-1.96$\pm$0.68&-0.90&0.10&-0.67\\
$\delta-L_{\rm c}$&\nodata&\nodata&\nodata&\nodata&\nodata&39.4$\pm$2.2&6.20$\pm$1.70&0.93&0.07&0.98\\
$\gamma_{\rm b}-L_{\rm c}$&39.7$\pm$1.0&0.86$\pm$0.20&0.95&0.05&0.95&40.8$\pm$1.4&2.55$\pm$0.54&0.96&0.04&0.96\\
$\gamma_{\rm b}-L^{'}_{\rm c}$&30.0$\pm$3.1&1.69$\pm$0.59&0.90&0.10&0.90&39.2$\pm$1.0&1.19$\pm$0.37&0.92&0.08&0.88\\
$B-\gamma_{\rm b}$&\nodata&\nodata&\nodata&\nodata&\nodata&3.6$\pm$0.1&-1.66$\pm$0.19&-0.99&0.01&-0.91
\enddata
\tablenotetext{1}{The correlation coefficient $r_{\rm bootstrap}$ is obtained by the bootstrap method as
described in the text.}
\end{deluxetable}

\label{lastpage}

\end{document}